\documentclass[]{article}
\usepackage{subeqnarray}
\newcommand{\mpl}{m_{Pl}}
\newcommand{\Omp}{{\cal O}(\mpl)}
\begin{document}

\begin{titlepage}

     \title{Condensation of Planckian Modes and the Inflaton}
\author{ Robert Brout\thanks{E-mail : physth@ulb.ac.be}\\
Service de Physique Th\'eorique, CP 225   \\
Universit\'e Libre de Bruxelles\\
Blvd du Triomphe\\
1050 Brussels - Belgium}
\date{\ }
\maketitle
\end{titlepage}
%\setlength{\baselineskip}{8mm}%Activer cette ligne et modifier
%la dimension pour changer l'interligne
\section*{Abstract}
To confront the transplanckian problem encountered in the
backward extrapolation of the cosmological expansion of
the momenta of the modes of quantum field theory, it is
proposed that there is a reservoir, depository of
transplanckian degrees of freedom.  These are solicited
by the cisplanckian modes so as to keep their density
fixed and the total energy density of vacuum at a
minimum.
The mechanism is due to mode - reservoir interaction,
whereupon virtual quantum processes give rise to an
effective mode-mode attraction. A BCS condensate results.
It has a massless and massy collective excitation, the
latter identified with the inflaton.  For an effective
non dimensional mode-reservoir coupling constant, $g\simeq 0.3$, the
order of magnitude of its mass is what is required to
account for cosmological fluctuations i.e. ${\cal
O}(10^{-6}\mbox{\rm{--}}10^{-5})m_{Planck}$.
\newpage
The establishment of the inflaton field scenario as the
mechanism of inflation must be counted as one of the
outstanding successes of physics in recent years \cite{1,2}.
Whereas the classical inflation was originally introduced \cite{3,4}
to implement the slow roll mechanism of inflation thereby
eliminating the flaws of the earlier inflationary
proposals \cite{5} which were introduced to provide for a causal
big bang and to explain flatness, it is now clear that
the inflaton must be regarded as a quantum field in order
to give a correct quantitative account of cosmological
fluctuations \cite{1,2}.

Yet, this quantum field seems to play no r\^ole in physics
other than to implement inflation.
Its mass, ${\mu}$, is estimated to be
$(10^{-6}\mbox{\rm{--}}10^{-5})m_{Pl}$ \cite{1} ($m_{Pl}$=Planck mass), so
if it does play any r\^ole in particle physics it will have to be in the
context of one of the conjectured hypotheses current in the field e.g.
grand unification or supersymmetry.

In this essay, we shall exhibit an
alternative possibility wherein the inflaton emerges as a collective
excitation of quantum gravity.  Its mass differs from the natural
scale, $\mpl$, by a sensitive reduction factor so that it is
possible to have $\mu << \mpl$ without undue tuning of parameters.

It is commonly admitted that quantum field theory (QFT),
expressed in terms of modes in interaction, becomes
inoperative for $|\vec{k}|>\Omp$; $k =$  momentum of the mode.  For
beyond the planckian scale, gravitation dominates.
In this we adhere to a non covariant Hamiltonian
formalism wherein $\vec{k}$ is the eigenvalue of the translation
operator on a standard space-like homogeneous surface of
our evolving universe i.e. our frame is our universe (We
hope in the future to develop a covariant formulation).

The difficulty is that $|\vec{k}|$ scales like $a^{-1}$ ($a$ = scale
factor).
Na\"{\i}vely, this would imply a reduction of the cut-off
proportional to $a^{-1}$, say ${\cal O}(10^{-50})$ from the beginning of
inflation to the present.  A planckian cut-off today
would place us inadmissably into the transplanckian
region in the past.

Furthermore, one believes that microscopic physics
should be but slightly influenced by the macroscopic
expansion, there being a factor of $10^{40}$ between the
present Hubble radius and that of the proton.  The
physics of matter is not conformally invariant.
Therefore QFT in the past should not differ sensibly from
QFT in the present.  Nor should the cut-off.

To accomplish this, we propose that the degrees of
freedom of fields fall into two classes, the cisplanckian
modes of QFT and the infinite remainder comprising a
transplanckian reservoir.  We may think of two fluids.
They are in interaction in such manner that the QFT modes
are solicited from the reservoir during the expansion so
as to keep their density (i.e. the cut-off) fixed.  In
this way the total energy density of vacuum is maintained
at its minimum, as befits the vacuum state.

The concept of a transplanckian reservoir may seem
unconventional, but it is in fact familiar, for example
in string theory.  Here the string degrees of freedom in
the massless sector, cut off at $\mpl$, comprise the stuff of
usual QFT, and the infinite remainder are in the transplanckian
modes, presumed to take up black hole configurations \cite{6},
perhaps modeling Wheeler's foam.

In this essay, however, we do not adopt a specific model of
the reservoir, but treat it as a phenomenological black box.
This is sufficient to deliver an inflaton and a formula
for its mass, essentially in terms of one dimensionless
parameter, (in addition to its natural scale, $\mpl$) in
attendance of a quantitative realization of a precise
model.

The very existence of a stable vacuum points to the
existence of a transplanckian reservoir.
Indeed, since the zero point energy density of QFT in the
presence of a cut-off scales like $a^{-4}$, the fluid of modes
will exert a "zero point pressure" resulting in its
infinite dilution in the absence of a compensating
"internal pressure".  Therefore there must be an
attractive interaction between modes and the reservoir.
(This consideration will ultimately lead to some
understanding of the cosmological constant, $\Lambda$, but that
is not the content of this paper).

The mode-reservoir interaction gives rise to mode-mode
attraction through the intermediary of the reservoir in
virtual processes (analog of phonon exchange among
superconducting electrons).  The consequent Cooper pair
formation leads to a BCS condensate \cite{7}.  Its massy,
collective excitation \cite{8} is identified with the inflaton.

We now draw up a simple mathematical expression of these
ideas.  We focus on mode-reservoir interactions since it
is these that lead to mode-mode attraction, whence a
condensate and the inflaton.  Usual QFT, which does not
allow for variation of mode number, is not available for
this task.

The interactions are taken to be of two sorts :
\begin{subeqnarray}
&&\rho_{-\vec{q}}\
a^{\dagger}_{\vec{k}+\vec{q}}\ a^{\dagger}_{-\vec{k}}+h.c.
\slabel{1a}\\
&&\rho_{-\vec{q}}\ a^{\dagger}_{\vec{k}+\vec{q}}\ a_{\vec{k}}+h.c.
\label{1b} \slabel{1b}
\end{subeqnarray}
$ \rho_{\vec{q}}$ is the $q^{th}$ Fourier transform of the reservoir
density. Operators $a_{\vec{k}}\ (a_{\vec{k}}^{\dagger})$ annihilate
(create) modes $\vec{k}$.  They have the properties $a_{\vec{k}}^2 = 0$ ;
$n_{\vec{k}}^2=n_{\vec{k}}$ ( where $n_{\vec{k}}=
a_{\vec{k}}^{\dagger}\ a_{\vec{k}} $ ) whence its eigenvalues are
$0$, $1$.  This is required because a mode either is or is not. There is
no need in what follows to specify commutation rules of
$a_{\vec{k}}$ for different $\vec{k}$'s.

Modes have zero point energy $(=\frac 12 \sum n_{\vec{k}}|\vec{k}|)$
where, for simplicity, the energy of mode $\vec{k}$ is taken equal to
$|\vec{k}|$.  This must be cut off, which we take to be at $|\vec{k}|
=K+(\Delta/2)$.  $K$ would be the cut-off of the uncondensed state,
a ``Fermi sea'' where $<n_{\vec{k}}>_{sea}=\theta(K-|\vec{k}|)$. Modes
relevant to condensation are denoted $\vec{k}\in \Delta$.  They lie within
a spherical shell of thickness $\Delta$ about $K$ where $K={\cal
{O}}(\mpl)$ and $\Delta$ is a finite fraction thereof of ${\cal{O}}(1)$.
$\Delta$ is a measure of the inverse response time of the reservoir when
it is perturbed by a mode through the action of Eqs (\ref{1a},\ref{1b}),
\cite{9}.

Up to an irrelevant subtraction the part of the kinetic
energy relevant to condensation is conveniently written
as
\begin{equation}
T_{red}=\sum_{\vec{k}\in
\Delta}\tilde\varepsilon_{\vec{k}}\ [n_{\vec{k}}+n_{-\vec{k}}-1]
\end{equation}
where $\tilde\varepsilon_{\vec{k}}=\frac14(|\vec{k}|-K)$.

The interactions,eqs (1), are responsible for several
phenomena.  Eq. (\ref{1a}) gives long lasting dissipative effects,
those we have invoked to maintain the 2-fluid equilibrium
of vacuum.
Eq. (\ref{1b}) contains the sense of a static interaction energy (the
term $\vec{q}=0$ in lowest order), as well as scattering of modes
by the reservoir.  We shall suppose that the sum of these
effects delivers a good approximation to the the true
vacuum.
This approximate vacuum has a filled Fermi sea of modes
for $|\vec{k}|<K$.  We call this unperturbed vacuum.  Without mode-
mode attraction induced by virtual effects such a vacuum
will not be a condensate and will not possess the
collective excitation we are seeking.  It is the analog
of a normal metal and not a superconductor.

To obtain a BCS condensate it is essential that the
induced mode-mode interaction be attractive.  The
mechanism for this attraction is that the perturbed
energy induced by the interactions, eqs (1), is due to
virtual transitions from the unperturbed ground state to
unperturbed excited states, hence a negative energy
denominator in second order.  Expressed more physically,
the presence of a mode polarizes the reservoir fluid
towards the space-time region where that mode is localized.
(See the above discussion of $\Lambda$).  A second mode is in
turn attracted to this region \cite{9}.

The attraction causes Cooper pair formation, hence an
instability of the unperturbed vacuum.  The BCS
condensate is built out of zero momentum pairs.  It is
constructed from the reduced hamiltonian \cite{7}.
\begin{equation}
H_{red}=T_{red}+\frac
12\sum_{\vec{k},\vec{k}'\in\Delta}V_{\vec{k},\vec{k}'}
b_{\vec{k}}b_{\vec{k}'}^{\dagger}\label{2}
\end{equation}
where $b_{\vec{k}}=a_{\vec{k}}a_{-\vec{k}}$ and $ T_{red}$ by eq. \cite{2}.
For $V_{\vec{k},\vec{k}'}<0$, the condensate is formed characterized 
by $<b_{\vec{k}}>_{cond}
\neq 0$ in
a small interval, $\mu$, about $K$.  The mass, $\mu$, obeys an
eigenvalue condition which in the approximation of a
momentum independent $V_{\vec{k},\vec{k}'}$ (simplification which does no
injustice to the physics here displayed)
\begin{equation}
1=|V|\sum_{\vec{k}\in 
\Delta}\frac{1}{\sqrt{\tilde\varepsilon^2_{\vec{k}}+\mu^2}}\simeq N\ 
|V|\ln
\frac{\Delta}{\mu}
\end{equation}
(for $\mu<<\Delta$). $N$ is the density of energies, 
$\tilde\varepsilon$, in $\Delta$.
As a result $\mu$ is of the form
\begin{equation}
\mu=\Delta\ \exp [-1/g^2]\label{5}
\end{equation}
where the non dimensional constant $g^2$ is a product of the
square of coupling constants multiplying $1$, the square
of the reservoir field, $\rho$, an $|\mbox{\rm{ energy denominator}}|$ and
$N$.

There are two collective excitations of the condensate \cite{8},
the massless gauge symmetry restorer (gauge symmetry
being broken by $<b_k>_{cond}\neq 0$) and a scalar whose mass is $\mu$.  This
is our inflaton.

To validate its candidature the inflaton must enjoy
certain properties.  Firstly it must be possible to build
up a classical field of large amplitude.  From experience
of spin wave theory this should not offer difficulties.
Less evident is how to characterize the mechanism(s) for
its conversion to the on-mass shell quanta of QFT (post-
inflationary heating).  This could involve its coupling
to the mode pairs out of which it is comprised and thence
to their excitation.  Or it may entail the intervention
of the reservoir.  One must investigate precise models to
elucidate these points, and to calculate $\mu$.

The main point of this essay has been to show that a
constant planckian cut-off of modes requires an elaborate
mechanism for its realization.  This in turn leads to a
BCS condensate and an inflaton.  For values of the
coupling constant $g(\simeq 0.3)$, one comes upon the required $\mu \simeq
(10^{-6}\mbox{\rm{--}}10^{-5})\mpl$. This obviates the necessity of 
extremely fine tuned
parameters offering some grounds for encouragement.

\section*{Acknowledgement}
I am deeply indebted to Renaud Parentani for many
discussions concerning these matters.

\end{document}